# Dielectric relaxation and predominance of NSPT and OLPT conduction processes in Ba$_{0.9}$Sr$_{0.1}$TiO$_3$

H. Zaitouni[1], L. Hajji[1*], E. Choukri[1], D. Mezzane[1], Z. Abkhar[1], L. Essaleh[1], A. Alimoussa[1],

M. El Marssi[2], I. A. Luk'yanchuk[2]

[1] Laboratory of Condensed Matter and Nanostructures (LMCN), Cadi-Ayyad University, Faculty of Sciences and Technology, Department of Applied Physics, Marrakech, Morocco.

[2] Laboratoire de physique de la matière Condensée (LPMC), Université de Picardie Jules Verne, 33 rue Saint-Leu, 80039 Amiens Cédex, France.

**Abstract**

We investigate the relaxation and conduction mechanism of Ba$_{0.90}$Sr$_{0.10}$TiO$_3$ (BST) ceramic, synthesized by the solid state reaction method. The dielectric and relaxation properties are analyzed in the temperature range of 380-450°C with alternative current in the frequency range of 20Hz-1MHz. Variation of dielectric constant, $\varepsilon'$, with temperature shows a normal ferroelectric transition at T$_c$=95°C with a weak degree of diffuseness. The modified Cole-Cole equation is used to describe all contributions to the relaxation mechanism. The frequency exponent $m(\omega,T)$ deduced from experimental data of the dielectric loss ($\varepsilon''$) as $m(\omega,T) = \left(\partial \ln \varepsilon'' / \partial \ln \omega \right)_T$ shows a temperature and frequency dependence. Two conduction process are observed: non-overlapping small-polaron tunneling (NSPT) at low frequencies and overlapping large polaron tunneling (OLPT) at high frequencies. The analysis of Nyquist plots reveals also the presence of two contributions, who which the activation energies have been calculated.

[*] Corresponding author: Pr. Lahoucine Hajji

Phone: (+212) 6 68 88 55 54,  E-mail: l.hajji@uca.ma; hajji1966@gmail.com

## 1. Introduction

Ferroelectric materials with perovskite structures have received much attention due to their remarkable functional properties, such as piezoelectricity, pyroelectricity, electro-optic effects, useful for microelectronic devices [1-3]. Among these systems, perovskite Barium Strontium Titanate, with general formula $Ba_xSr_{1-x}TiO_3$ (BST), is an attractive candidate for applications in decoupling capacitors, storage capacitors, and dielectric field tunable elements for high-frequency devices [4,5,6].

The microstructure and dielectric properties of BST both in bulk and in thin films forms have been reported by several authors [7,8]. However, the literature survey shows limited reports on the relaxation processes and conduction mechanism in BST system, although the understanding of these features is very important to realize the practical applications. Considering that the solid defects play a decisive role in all of these applications, it is very important to gain a fundamental understanding of the conductive mechanism. Various kinds of defects, for instance, are oftenly suggested as being responsible for the dielectric relaxations at high temperature range.

In the present work, we focused on the relaxation processes and conduction mechanism of $Ba_{0.90}Sr_{0.10}TiO_3$ (BST) ceramic, synthesized by the solid state reaction method. The modified Cole-Cole equation is used to describe all contributions to the relaxation mechanism in BST and the activation energies are calculated for these contributions. In addition, using frequency and temperature dependence of the exponent *m*, we discuss the conduction mechanism and analyzed the experimental data. The combined effect on the complex impedance spectroscopy and modulus formalism has been employed to explore the electrical response of grain and grain boundaries.

## 2. Experimental procedure

$Ba_{0.9}Sr_{0.1}TiO_3$ (BST) ceramic was synthesized by conventional solid-state reaction method. The starting raw materials were $BaCO_3$, $SrCO_3$ and $TiO_2$ powders, which were weighed based on stochiometric composition of $Ba_{0.9}Sr_{0.1}TiO_3$. The weighed powders were mixed with ethanol in an agate mortar for 2h and then dried and calcined for 12 h at 1150°C. After calcination, 5wt % of polyvinyl alcohol (PVA) was added to the calcined powder as binder and the pellets having 1-2 mm thickness and 13 mm diameter were pressed using the uniaxial hydraulic press. The pressed pellets were burned out up to 700°C to remove PVA and then sintered at 1325°C. The permittivity was measured with an impedance meter HP 4284A connected to computer, in the frequency range from 20Hz to 1MHz. The powders calcinated were characterized by Xray diffraction (XRD). The lattice parameters were calculated and refined by the Full-prof program.

## 3. Theoretical models

One of the most important aspects in the dielectric response of ferroelectric materials is the dielectric relaxation phenomenon. The first empirical expression for the dielectric constant ε was given by Cole-Cole [9].

$$\varepsilon = \varepsilon_\infty + \frac{(\varepsilon_s - \varepsilon_\infty)}{1 + (i\omega\tau_0)^{1-\alpha}} \quad (1)$$

where $\varepsilon_s$ and $\varepsilon_\infty$ are the static and high frequency limits of dielectric permittivity, respectively, α is the Cole-Cole parameter (0<α<1) represent the degree of the disorder, and τ is the most probable relaxation time. This equation ignores the effect of the conductivity. However, the conductivity in ferroelectric cannot be neglected specially in low frequency and high temperature ranges. According to the Maxwell equation, the current density is given by

$$\vec{J} = \vec{J}_l + \vec{J} = \sigma\vec{E} + j\varepsilon_0\varepsilon\omega\vec{E} \quad (2)$$

and

$$\varepsilon^* = \varepsilon - j\sigma/\varepsilon_0\omega \quad (3)$$

where σ is the electric conductivity for free charges. To take into account the possible interaction of free charges with the lattice, σ must be replaced by the complex conductivity, $\sigma^* = \sigma_1 + j\sigma_2$ [10] in the complex dielectric permittivity, expressed as

$$\varepsilon^* = \varepsilon - j\sigma^*/\varepsilon_0\omega^s \quad (4)$$

the frequency exponent *s* is introduced to take into account the disorder effect in the conductivity.

From Eqs. (1) and (4), the real and the imaginary permittivity parts are

$$\varepsilon' = \varepsilon_\infty + \frac{(\varepsilon_s - \varepsilon_\infty)\{1 + (\omega\tau)^\beta \cos(\beta\pi/2)\}}{1 + 2(\omega\tau)^\beta \cos(\beta\pi/2) + (\omega\tau)^{2\beta}} + \sigma_2/\varepsilon_0\omega^s \quad (\beta = 1-\alpha) \quad (5)$$

$$\varepsilon'' = \frac{(\varepsilon_s - \varepsilon_\infty)\{(\omega\tau)^\beta \sin(\beta\pi/2)\}}{1 + 2(\omega\tau)^\beta \cos(\beta\pi/2) + (\omega\tau)^{2\beta}} + \sigma_1/\varepsilon_0\omega^s = \varepsilon''_d + \varepsilon''_c \qquad (6)$$

where $\varepsilon''_d$ is the contribution associated with the dielectric relaxation due to spaces charges and dipole orientation, and $\varepsilon''_c$ is the contribution associated with the mobile charges carriers.

In this work, the contribution of the polarization ($\varepsilon''_d$) and conductivity ($\varepsilon''_c$) to the dielectric relaxation are identified by fitting the experimental data of $\varepsilon''$ using Eq. (6).

According to Guintini's theory [11], $\varepsilon''$ at a given frequency is expressed as

$$\varepsilon'' = A\omega^m \qquad (7)$$

where $A$ depends only on temperature and $m$ is the frequency exponent which represent the disorder effect in the conductivity.

The ac conductivity can be calculated using the "universal dielectric response" (UDR)

$$\sigma_{ac}(\omega) = \sigma_{tot}(\omega) - \sigma_{dc} = B\omega^s \qquad (8)$$

where $B$ is pre-exponential factor and $s$ is the frequency exponent [12]. $\sigma_{ac}$ was calculated from imaginary part of the dielectric constant by

$$\sigma_{ac}(\omega) = \varepsilon_0 \omega \varepsilon''(\omega) \qquad (9)$$

Using Eqs. (7), (8) and (9), we obtained

$$s = 1 + m \qquad (10)$$

For our best knowledge; all the previous works used the simplest description where $m$ was considered as frequency independent. We show here that this hypothesis is slightly inaccurate. The parameters $m$ and $s$ that according to Eqs. (7) and (8) are determined as

$$m(\omega, T) = (\partial \ln \varepsilon''/\partial \ln \omega)_T \qquad (11)$$

$$s(\omega, T) = (\partial \ln \sigma_{ac}/\partial \ln \omega)_T \qquad (12)$$

may change considerably as function of frequency.

The complex electric modulus $M^*(\omega)$ was calculated using the following expression

$$M^*(\omega) = M' + jM'' = 1/\varepsilon^* \tag{13}$$

We identify the dominant conducting mechanism in a given range of frequency by comparing the behavior of *m* with the modulus *M″* as a function of frequency and temperature.

## 4. Results and discussion

### 4.1. Structure and dielectric studies

Structural properties of BST were determined by X-ray diffraction (XRD) patterns analysis. Fig. 1 shows the room temperature XRD pattern. All the peaks in the patterns are matching and showing purely tetragonal single phase crystal related to the tetragonal $BaTiO_3$. The analysis of the spectrum by the refinement program Full-Prof showed that our material crystallizes in tetragonal perovskite structure at room temperature with space group P4mm. The lattice parameters obtained are a=b=3.98 Å and c=4.01 Å. These values are close to a=b=3.97 Å and c=3.99 Å reported for BST [8].

**(Insert Fig. 1, here)**

The real and imaginary parts of dielectric constant, were determined using the following expressions

$$\varepsilon' = Cd/\varepsilon_0 A \tag{14}$$

$$\varepsilon'' = \varepsilon' tg\delta \tag{15}$$

where *d* is the sample thickness, *A* the area of the electrode, *C* the capacity and *tgδ* the dielectric loss.

The temperature dependence of both ε' and *tgδ* in the frequency range 500Hz-500KHz are shown in Fig. 2. The inset, represent the inverse of ε' as a function of temperature. The dielectric constant increases with increasing temperature and shows a peak at $T_C$ = 95°C, which is a characteristic feature of ferroelectric–paraelectric phase transition. This behavior is in good agreement with literature [13]. We note also that there is no frequency dispersion of permittivity above $T_C$, the dielectric constant follows the Curie-Weiss law [14] given by

$$\varepsilon' = \frac{\varepsilon'_m}{1 + \left(\frac{(T-T_m)}{\Delta}\right)^{\gamma}} \quad (16)$$

where $\varepsilon'_m$ is the dielectric permittivity maximum, $T_m$ temperature of the real part dielectric maximum, γ indicates the character of the paraelectric–ferroelectric phase transition and Δ is the peak broadening, which defines the diffuseness degree of the ferroelectric–paraelectric phase transition. Fig. 3 shows the good agreement between experimental data of ε' with those calculated by Eq. (16). The adjustable parameters $\varepsilon'_m, T_m, \gamma$ and Δ are listed in Table 1 for various frequencies. The obtained value of γ≈1 confirms the normal ferroelectric behaviors of BST with a diffuseness degree with Δ=12 that is lower than the reported value for BaTiO$_3$ [15].

**(Insert Fig. 2, here)**

**(Insert Fig. 3, here)**

**(Insert Table 1, here)**

We have demonstrated in Fig. 4 the variation of the dc-electrical conductivity as a function of frequency in the temperature range 380-440°C. A linear behavior of $\ln \sigma_{dc}$ vs. $(1/T)$ is observed with an activation energy $E_a$ deducted by the expression

$$\sigma_{dc} = \sigma_0 \exp(-E_a/KT) \quad (17)$$

here $\sigma_0$ is the pre-exponential factor. We obtained $E_a = 1,3 eV$.

**(Insert Fig. 4, here)**

The frequency dependence of ε' and ε" for various temperatures are represented in Fig. 5 and Fig. 6, respectively.

We observe that both ε' and ε" decrease with frequency for all temperatures and increase with temperature. Good fitting of ε' and ε" is obtained using Eqs. (5) and (6).

**(Insert Fig. 5, here)**

**(Insert Fig. 6, here)**

We can distinguish two domains. The low frequency domain, where the contribution to ε" is governed by the free charges carries (giving by $\sigma_1$) and the contribution to ε' is governed by the charge carrier localized at defect sites and interfaces (giving by $\sigma_2$). At high frequency domain, the two contributions become insignificant due to the high frequency of the field which reduces the contribution of the charge carriers. The parameters used in the calculation of ε" are listed in Table 2.

**(Insert Table 2, here)**

Using the Eq. (6), we can deduce the dielectric relaxation component $(\varepsilon_d^")$ from ε" by subtracting the conduction component $(\varepsilon_c^")$. The frequency where $(\varepsilon_d^")$ reaches its maximum is corresponds to the relaxation frequency $(\omega_r)$. As temperature increases, $\omega_r$ shifts to the higher frequencies demonstrating a temperature dependence of relaxation time $(\tau = 1/\omega_r)$. This dependence which satisfies the Arrhenius law.

$$\tau_r = \tau_0 \exp(E_{relax}/KT) \qquad (18)$$

is shown in Fig. 10. Here $\tau_0$ is the pre-exponential factor and $E_{relax}$ is activation energy for dielectric relaxation. The relaxation parameters, $(E_{relax} = 1,27eV, \tau_0 = 3,3.10^{-12} s)$ were deduced from the fitting of $\ln \tau_r$ vs. $(1/T)$. These values are in the same range as those reported for oxygen diffusers [16].

In order to analyze the dynamic electrical conduction at low and high frequencies, we plot in Fig. 7, the variation of the relaxation $\ln(\varepsilon")$ versus $\ln(\omega)$ for the high temperature region. In the two domains (LF, HF), we observe a series of lines with different slopes. The frequency exponent $m(\omega,T)$ given by Eq. (11) is the slope of the curves $\ln(\varepsilon") = f(\ln(\omega))$. Fig. 8 reveals that this exponent is temperature and frequency dependent. In the same Fig. 8 we represent also the variation of the imaginary part of modulus with frequency for the same temperatures.

**(Insert Fig. 7, here)**

**(Insert Fig. 8, here)**

The correlation between $M"$ and $\varepsilon"$ (Fig. 8) shows that the behavior of the coefficient $m$ changes really with frequency and temperature. It is interesting to note that the frequency where we obtained a change in the behavior of $m$ with frequency and temperature coincides with the frequency where $M"$ reach its maximum. The frequency domain below $\omega_m$, where the exponent $m$ decreases with frequency and increases with temperature corresponds to the non-overlapping small-polaron tunneling theoretical model (NSPT) [17]. Another one from above $\omega_m$, where $m$ increases with frequency and decreases with temperature correspond to overlapping large polaron tunneling theoretical model (OLPT) [17].

We conclude that the crossover from NSPT, where polarons are localized, to OLPT, where polarons are delocalized, can be due to the effect of the increasing of the ac electrical conductivity with the increasing of the frequency.

4.2. Impedance spectroscopy

In order to determine the contributions (grain, grain boundaries and electrodes), we use the impedance spectroscopy. Fig. 9 illustrates the Nyquist plot (-$Z"$ versus $Z'$) in the temperature range 400-440°C. The radius of curvature was found to decrease with increasing temperature which shows the increase in conductivity of the sample with temperature. We observe that the experimental data are located on two semicircles that can be attributed to the bulk and grain boundary contributions at high and lower frequency, respectively. The complex impedance data have been analyzed using an equivalent two R//CPE circuits connected in series. The good conformity between experimental and calculated curves indicates that the proposed electric equivalent circuit describe well the behavior of our data. The obtained fitting parameters are shown in Table 3. The relaxation times for the grain and grain boundary were calculated using

$\tau_i = (R_i Q_i)^{\frac{1}{n_i}}$. These relaxation times obey to Arrhenius law with the activation energies 1.53 eV and 1.47 eV for grain and grain boundary, respectively as shown in Fig. 10. We note that these values are very close suggesting that the conduction in grain and grain boundary originate from the same physical mechanism.

**(Insert Fig. 9 here)**

**(Insert Fig. 10 here)**

**(Insert Table 3 here)**

We have also adopted the modulus formalism to study the relaxation mechanisms in BST. Fig. 11 shows the variation of $M''$ versus angular frequency at higher temperatures. These curves are characterized by the presence of the only one relaxation peak attributed to the grain contribution. The position of this peaks shifts to higher frequencies as the temperature increases suggesting that the dielectric relaxation is thermally activated. In order to calculate the activation energy of the grain relaxation, we have modeled our experimental data using the Kohlrausch Willians Watts (KWW) function proposed by Bergman [18]

$$M''(\omega) = \frac{M''_{max}}{(1-\beta) + \left(\frac{\beta}{1+\beta}\right)\left[\beta(\omega_{max}/\omega) + (\omega/\omega_{max})^\beta\right]} \quad (19)$$

The temperature dependence of the most probable relaxation time $\tau_m = 1/\omega_m$ obeys the Arrhenius behavior with activation energy of relaxation 1.53 eV in the vicinity of the activation energy of conduction (Fig. 10). We can deduce that the conduction and the relaxation process are ascribed by the same mechanism. The obtained parameter β= 0.6 indicates on the deviation from the Debye-type relaxation.

**(Insert Fig. 11 here)**

## 5. Conclusion

Ba$_{0.90}$Sr$_{0.10}$TiO$_3$ was synthesized by the solid state reaction method. The phase formation is confirmed by XRD technique. Rietveld analysis of BST shows a good agreement between the experimental and theoretical line profile. A modified Cole-Cole equation has been used to study dielectric relaxation in this material. The frequency and temperature dependence of the exponent *m* has been theoretically supported by the NSPT model at low frequency and OLPT model at high frequency. The same activation energy obtained by dc conductivity, impedance and modulus formalism shows that all dynamic processes may be attributed to the same type of charge carriers.

**Table captions**

Table 1: The parameters $\varepsilon'_m$, $T_m$, $\Delta$ and $\gamma$ for BST at several frequencies.

Table 2: Fitted values of various parameters for Eq. 6.

Table 3: The extract parameters for the circuit elements.

**Figure captions**

Figure 1: Observed and calculated X-ray diffraction pattern of BST sample.

Figure 2: Dielectric permittivity and dielectric loss of BST. Inset shows the Curie-Weiss law plot.

Figure 3: Theoretical and experimental curves of ε′ vs. $T$ at different frequencies.

Figure 4: Temperature dependence of dc-electrical conductivity.

Figure 5: Theoretical and experimental curves of ε' vs. angular frequency at several temperatures. The inset shows only the dielectric contribution.

Figure 6: Theoretical and experimental curves of ε" vs. angular frequency at several temperatures. The inset shows only the contribution of the dielectric relaxation.

Figure 7: Variation of the dielectric relaxation $\ln(\varepsilon")$ versus $\ln(\omega)$ at several temperatures.

Figure 8: Variation of the modulus $M″$ and the coefficient $m$ with frequency at several temperatures.

Figure 9: Nyquist plots of the BST at different temperatures. The fitted curve is shown in the inset for a representative temperature of 440°C.

Figure 10: Variation of the relaxation time with inverse of temperature for BST.

Figure 11: Frequency dependence of $M″$ at various temperatures.

Table 1

| Frequency | $\varepsilon'_m$ | $T_m$ (°C) | Δ (°C) | γ |
|---|---|---|---|---|
| 500 Hz | 14771 | 95,76 | 12,63 | 1,03 |
| 1 KHz | 14735 | 95,75 | 12,35 | 1,01 |
| 10 KHz | 14376 | 95,75 | 12,45 | 1,01 |
| 100 KHz | 14095 | 95,73 | 12,55 | 1,02 |
| 500 KHz | 14628 | 95,73 | 12,28 | 1,04 |

Table 2

| T(°C) | $\Delta\varepsilon = \varepsilon_s - \varepsilon_\infty$ | τ(s) | β | $\sigma_1$ (Sm$^{-1}$) | s |
|---|---|---|---|---|---|
| 400 | 13784 | 7,43.10$^{-4}$ | 0,71 | 1,43.10$^{-5}$ | 0,82 |
| 410 | 14092 | 5,26.10$^{-4}$ | 0,71 | 2,18.10$^{-5}$ | 0,84 |
| 420 | 14351 | 3,79.10$^{-4}$ | 0,71 | 3,31.10$^{-5}$ | 0,86 |
| 430 | 14271 | 2,80.10$^{-4}$ | 0,73 | 4,90.10$^{-5}$ | 0,87 |
| 440 | 14578 | 2,13.10$^{-4}$ | 0,73 | 8,04.10$^{-5}$ | 0,90 |

Table 3

| T(°C) | $R_g$ (Ω) | $Q_g$ (F) | $n_g$ | $\tau_g$ (s) | $R_{gb}$ (Ω) | $Q_{gb}$ (F) | $n_{gb}$ | $\tau_{gb}$ (s) |
|---|---|---|---|---|---|---|---|---|
| 400 | 17100 | 1,47.10$^{-9}$ | 0,957 | 1,55E-05 | 350790 | 7,14.10$^{-8}$ | 0,746 | 7,13.10$^{-3}$ |
| 410 | 11337 | 1,49.10$^{-9}$ | 0,955 | 1,01E-05 | 254150 | 8,52.10$^{-8}$ | 0,733 | 5,36.10$^{-3}$ |
| 420 | 7621 | 1,45.10$^{-9}$ | 0,957 | 6,61E-06 | 179930 | 1,02.10$^{-7}$ | 0,721 | 3,91.10$^{-3}$ |
| 430 | 5844 | 1,70.10$^{-9}$ | 0,943 | 4,94E-06 | 126580 | 1,09.10$^{-7}$ | 0,723 | 2,68.10$^{-3}$ |
| 440 | 4318 | 1,79.10$^{-9}$ | 0,937 | 3,50E-06 | 82447 | 1,23.10$^{-7}$ | 0,719 | 1,69.10$^{-3}$ |

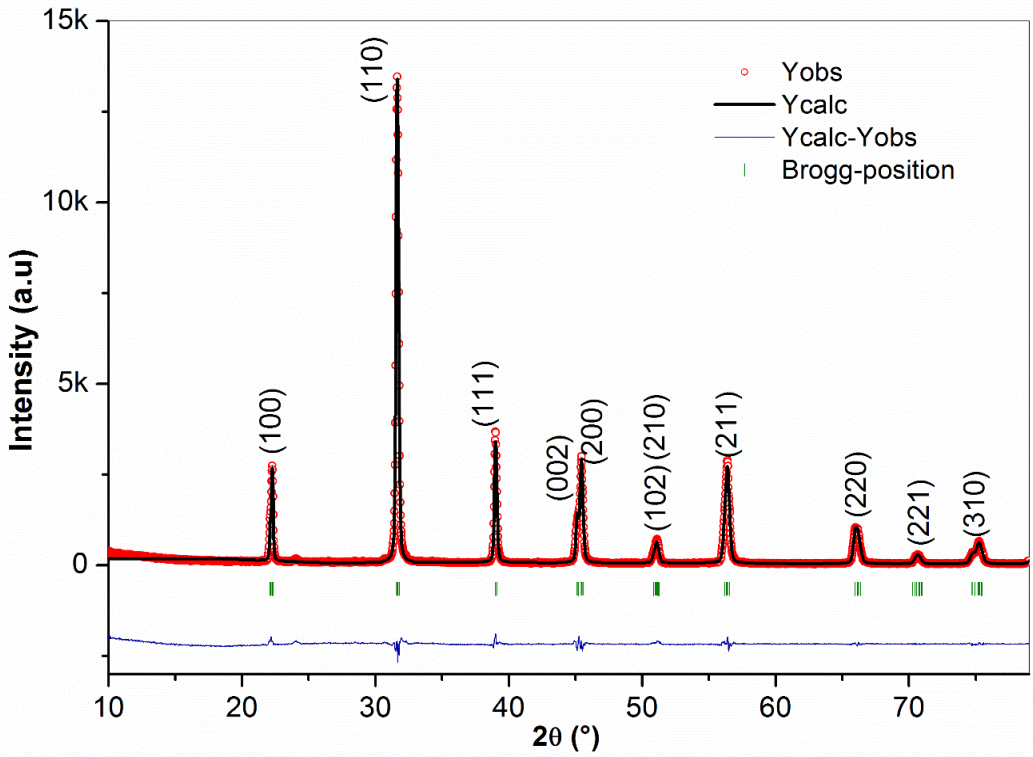

Figure 1

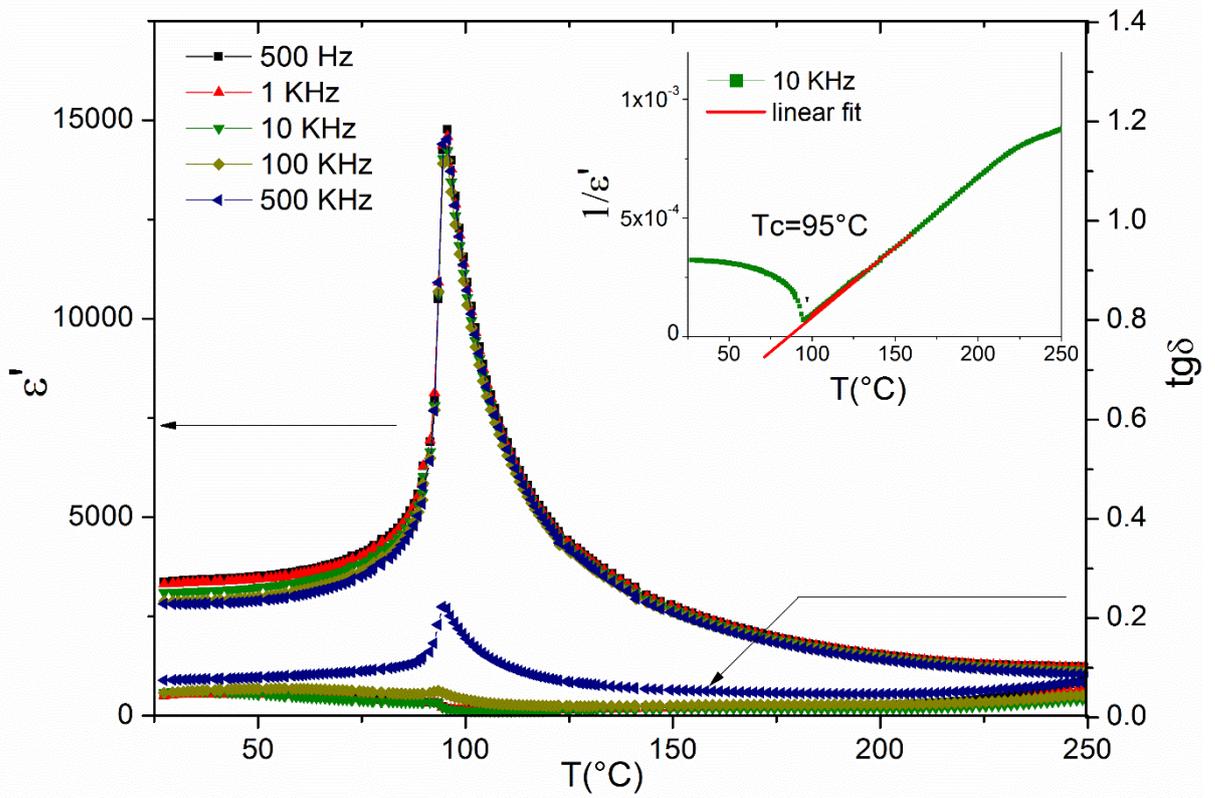

Figure 2

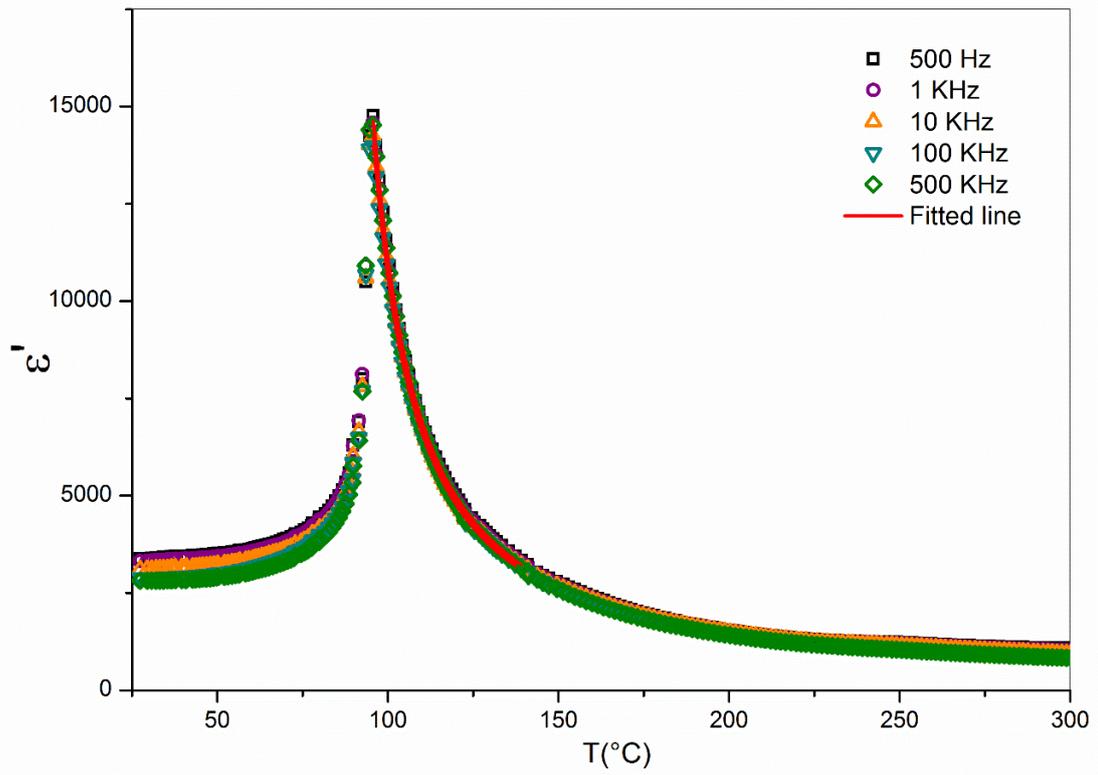

Figure 3

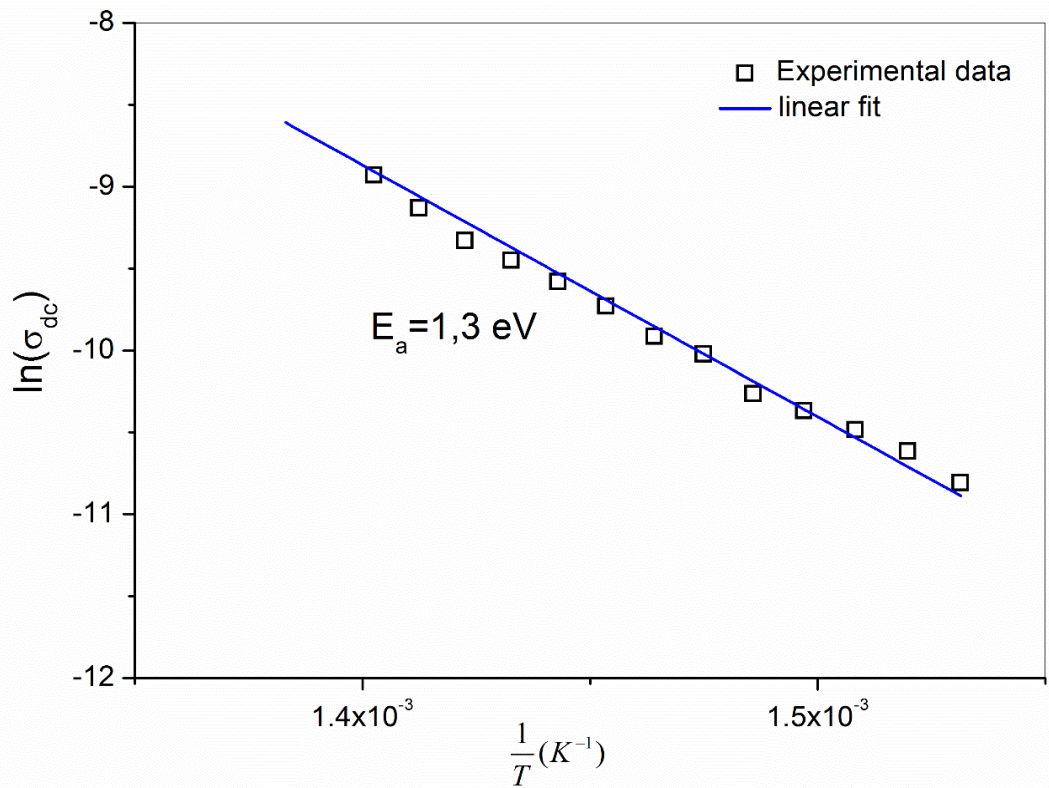

Figure 4

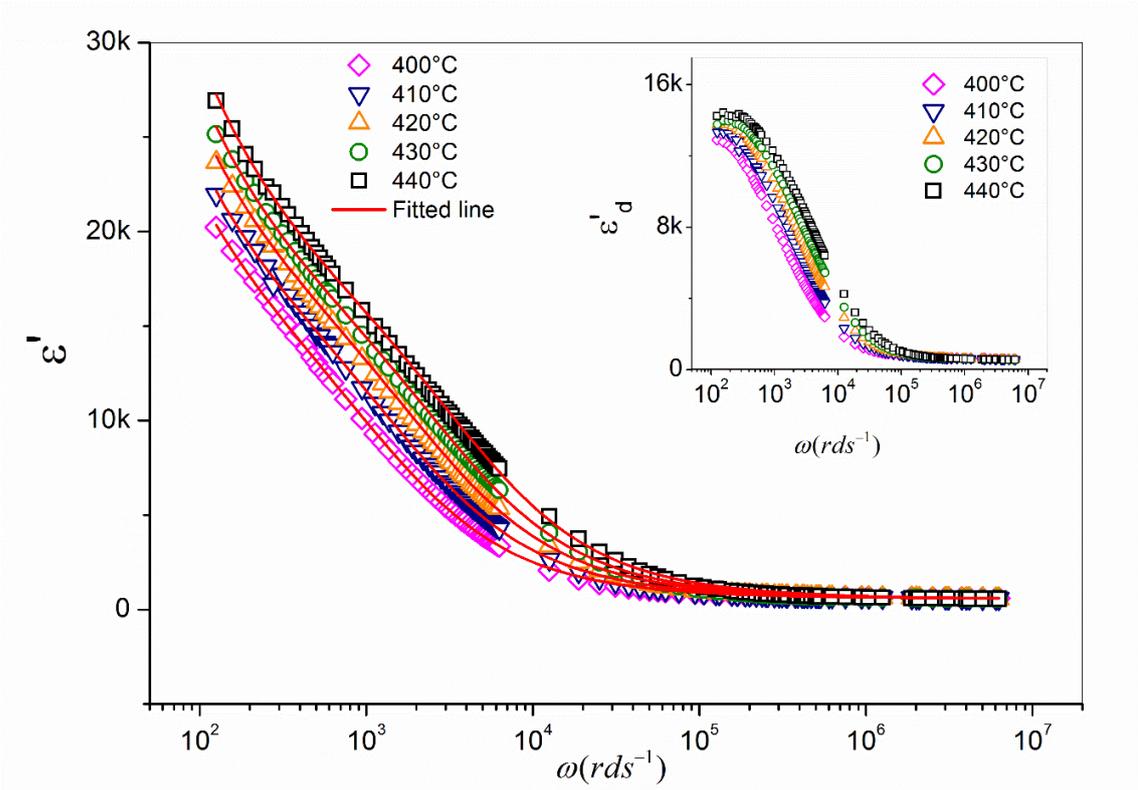

Figure 5

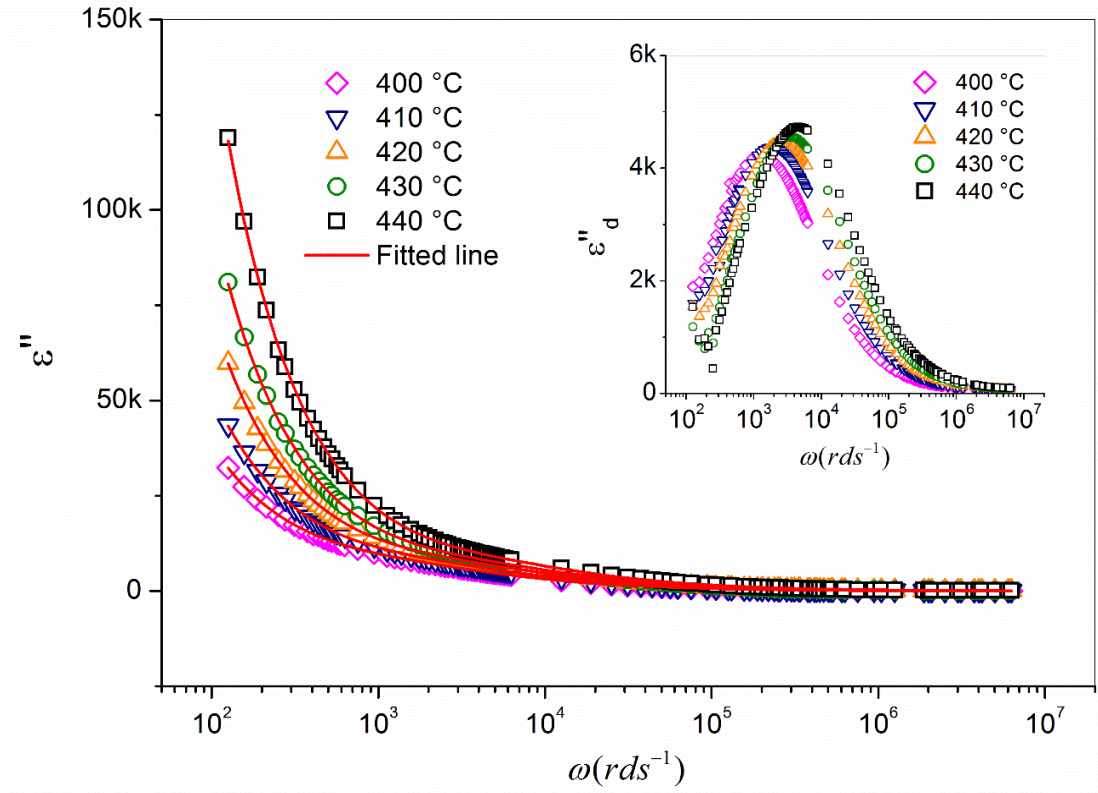

Figure 6

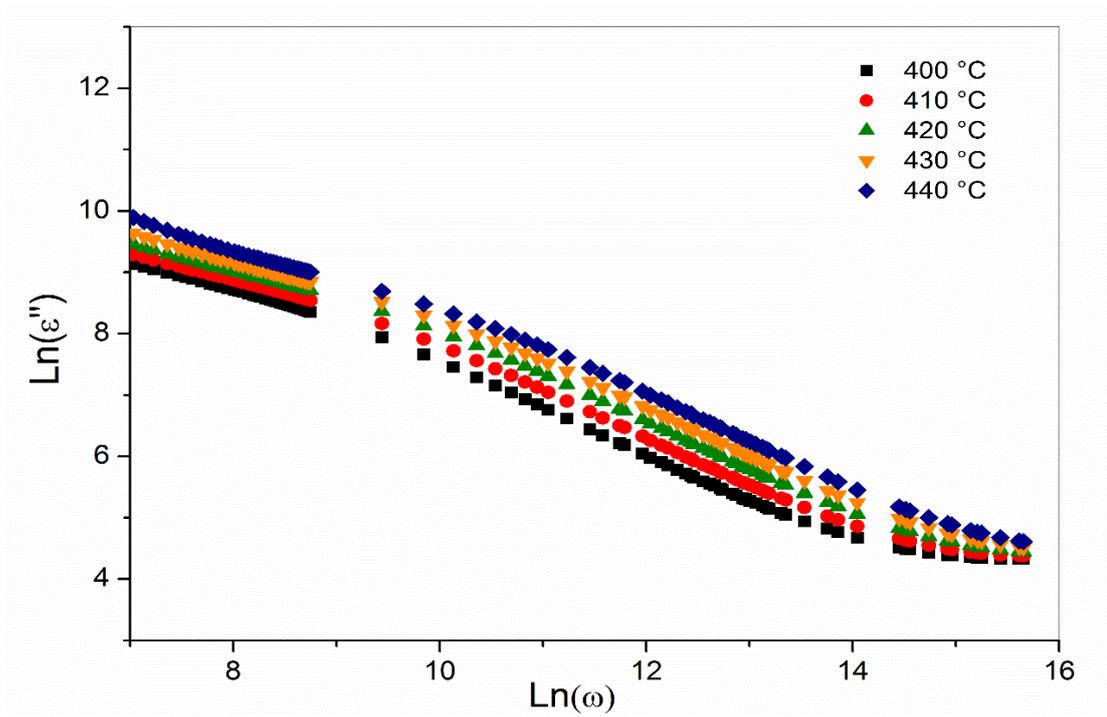

Figure 7

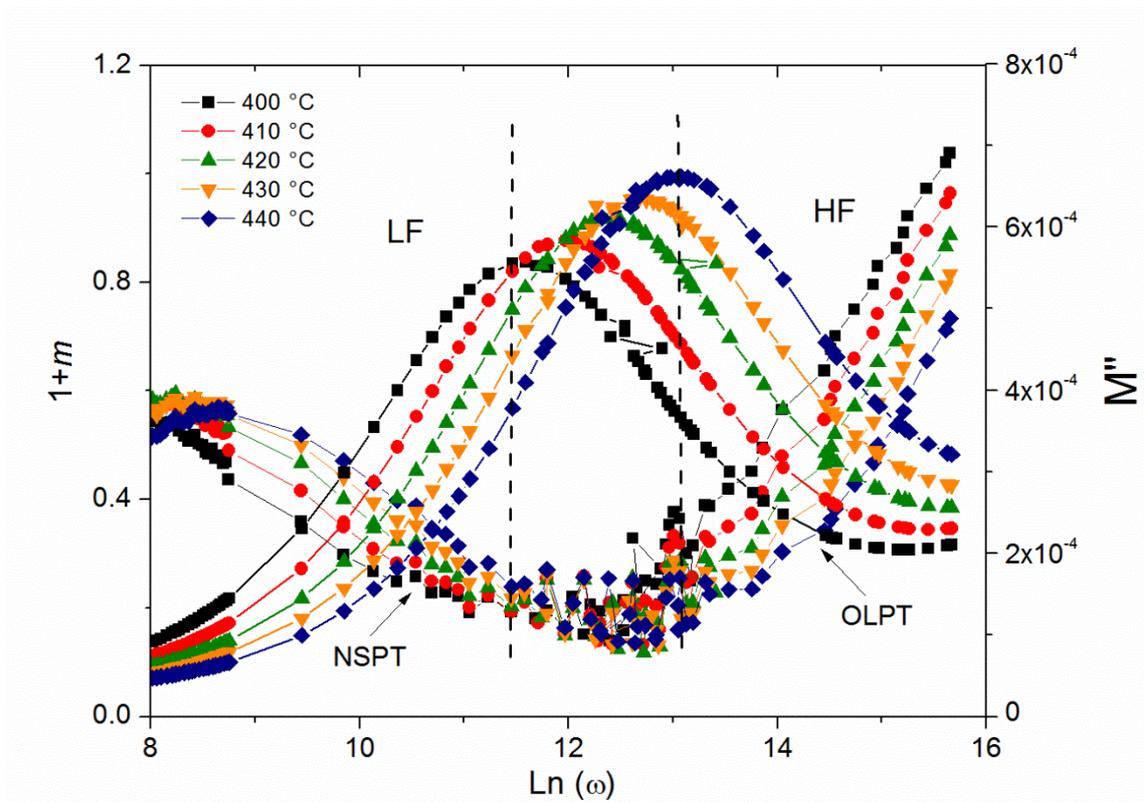

Figure 8

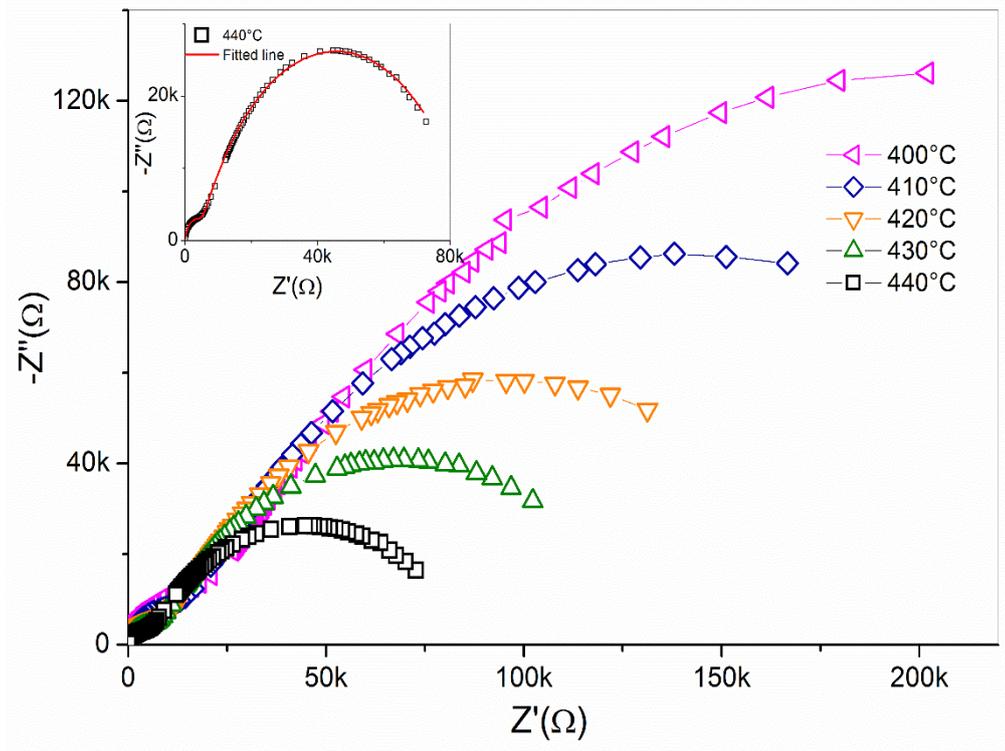

Figure 9

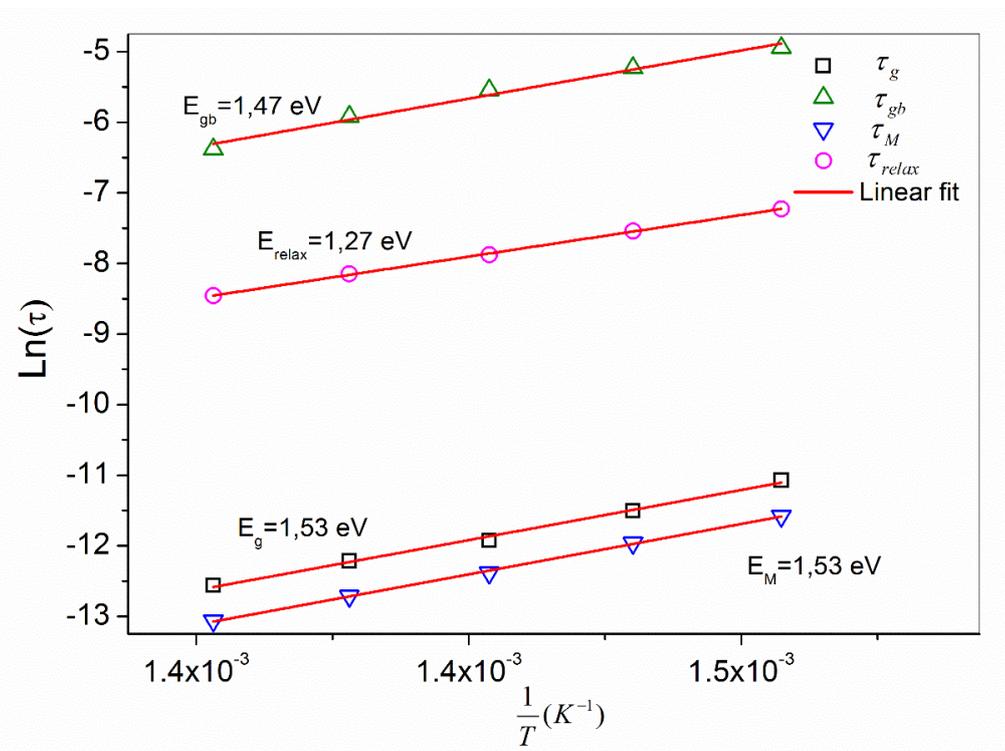

Figure 10

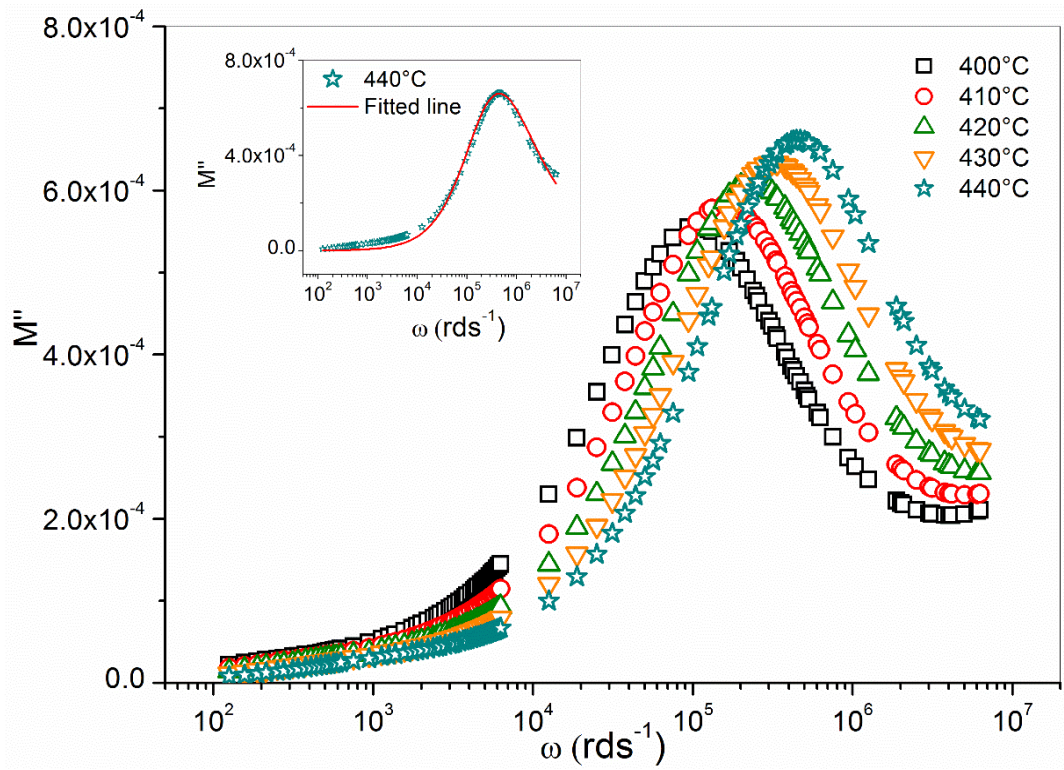

Figure 11